\documentstyle[12pt]{article}
\begin{document}
\begin{center}
{\large\bf ON THE ROLE OF FSI IN $K \to 2\pi$ DECAY}\\
\vspace{5mm}
{\large E.Shabalin} \\ Institute for Theoretical and Experimental Physics,
B.Cheremushkinskaya 25, 117218 Moscow, Russia \\
\end{center}
\vspace{1cm}
\begin{abstract}
Contrary to wide-spread opinion that the final state interaction (FSI)
enlarges the amplitude $<2\pi;I=0|K^0>$, we argue that FSI is not able
to increase the absolute value of this amplitude.
\end{abstract}
\vspace{1cm}
\section{Introduction}
The great progress in understanding the nature of the $\Delta I=1/2$
rule in $K\to 2\pi$ decays was achieved in the paper \cite{1}, where
the authors had found a considerable enlargement of contribution of the
operators containing a product of left-handed and right-handed quark
currents generated by the diagrams called later the penguin ones.  But for
a quantitive agreemement with the experimental data, a search for some
additional enlargement produced by long-distance effects was utterly
desirable.

In the literature, two possible mechanisms of such increase were discussed.
First one was based on assumption that the strenthening of $s$-wave $2\pi$
amplitude with isospin $I=0$ arises due to small mass of the
intermediate scalar $\sigma$ meson. And the calculations in the framework
of chiral theory to leading order in momentum expansion of $<2\pi;I=0|K>$
amplitude confirmed such a possibility \cite{2}, \cite{3}.

The second mechanism of strengthening the $<2\pi;I=0|K>$ amplitude was
ascribed to final state interaction of the pions \cite{4}-\cite{12}.

In the papers \cite{4}-\cite{7}, the calculations were
based on the dispersion relation
\begin{equation}
\mbox{Re}A^I_j(s)=\frac{1}{\pi}\int_{4\mu^2}^{\infty} dz
\frac{\mbox{Im}A^I_j(z)}{z-s+i\varepsilon} +\mbox{subtractions}
\end{equation}
But $\mbox{Im}A^I_j(z)$ itself is unknown at $z>m^2_K$, the channels
$K \bar K$, $4\pi$ and so on contribute at the values $z$ larger than
threshold of elasticity and, in addition, even at $z$ smaller than
threshold of elasticity, the resonances of the type of $\sigma$ mesons are
possible \cite{13} changing  considerably the expression for
$\mbox{Im}A^I_j(z)$ calculated in the framework of ChPT and used in
eq.(1).

All these circumstances are the cause to reexamine the role of FSI,
using some other approach. It will be done in present paper basing on
linear $U(3)_L\bigotimes U(3)_R$ $\sigma$ model with broken symmetry.

\section{The technique of
calculation} Our calculations are based on employment of the effective
lagrangian of non-leptonic, strangeness-changing weak interaction
represented in \cite{1} \begin{equation} L^{weak} =\sqrt{2}G_F \sin
\theta_C \cos \theta_C \sum_i c_i O_i, \end{equation} where
\begin{equation}
\begin{array}{llllll}

O_1=\bar s_l \gamma_{\mu}d_L \cdot\bar u_L \gamma_{\mu}u_L-\bar s_L
\gamma_{\mu}u_L \cdot \bar u_L\gamma_{\mu}d_L \qquad(\{8_f\},\Delta I=1/2),
\\
O_2=\bar s_L \gamma_{\mu}d_L \cdot
 \bar u \gamma_{\mu} u_L+\bar s_L \gamma_{\mu}u_L \cdot \bar u_L
\gamma_{\mu} d_L + 2 \bar s_L \gamma_{\mu} d_L \cdot \bar d_L \gamma_{\mu}
d_L   \\ +2\bar s_L \gamma_{\mu}d_L \cdot \bar s_L \gamma_{\mu}s_L \qquad
(\{8_d\}, \Delta I=1/2)  \\ 
O_3=\bar s_L \gamma_{\mu}d_L \cdot \bar u_L \gamma_{\mu}u_L +\bar s_L
\gamma_{\mu}u_L \cdot \bar u_L \gamma_{\mu} d_L +2\bar s_L \gamma_{\mu}
d_L \cdot \bar d_L \gamma_{\mu}d_L \\
-3 \bar s_L \gamma_{\mu}d_L \cdot \bar s_L \gamma_{\mu}s_L \qquad
(\{27\}, \Delta I=1/2) \\
O_4= \bar s_L \gamma_{\mu}d_L \cdot \bar u \gamma_{\mu}u_L +
\bar s_L \gamma_{\mu}u_L \cdot \bar u_L \gamma_{\mu}d_L \\
-\bar s_L \gamma_{\mu}d_L \cdot \bar d_L \gamma_{\mu}d_L, \qquad
(\{27\}, \Delta I=3/2),  \\
O_5=\bar s_L\gamma_{\mu} \lambda^a d_L(\sum_{q=u,d,s} \bar q_R
\gamma_{\mu} \lambda^a q_R) \qquad (\{8\}, \Delta I=1/2), \\
O_6=\bar s_L \gamma_{\mu} d_L(\sum_{q=u,d,s} \bar q_R \gamma_{\mu} q_R),
\qquad (\{8\}, \Delta I =1/2).
\end{array}
\end{equation}
This set is sufficient for calculation of the CP-even part of  $K \to
2\pi$ amplitude.

The bosonization of the operators $O_i$ in linear $U(3)_L \bigotimes
U(3)_R$ $\sigma$ model is carried out using the relations \begin{equation}
\begin{array}{lll}
\bar q_i \gamma_{\mu}(1+\gamma_5)
q_j=i\left[(\partial_{\mu}U)U^{\dag} -U(\partial_{\mu} U^{\dag}
)\right]_{ji} \\ \bar q_i(1+\gamma_5) q_j=-\frac{\sqrt{2}F_{\pi}
m^2_{\pi}}{m_u+m_d} U_{ji}, \qquad F_{\pi}=93 \; \mbox{MeV} \end{array}
\end{equation}
where
\begin{equation}
U=\hat \sigma +i\hat \pi, \qquad <\sigma_0+
\sigma_8/\sqrt{2}>_0=\sqrt{\frac{3}{2}} F_{\pi} \end{equation} and $ \hat
\sigma $ is $3\times 3 $ matrix of the scalar partners of the nonet of
pseudoscalar mesons:  \begin{equation} \left( \begin{array}{lll}
\frac{\pi_0}{\sqrt{3}}+\frac{\pi_8}{\sqrt{6}}+\frac{\pi_3}{\sqrt{2}}&
\pi^+ & K^+ \\
\pi^- &
\frac{\pi_0}{\sqrt{3}}+\frac{\pi_8}{\sqrt{6}}-\frac{\pi_3}{\sqrt{2}} & K^0
\\
K^- & \bar K^0  & \frac{\pi_0}{\sqrt{3}}-\frac{2\pi_8}{\sqrt{6}}
\end{array}
\right).
\end{equation}

The operators $O_{1-3}$ produce the following part of $<2\pi;I=0|K^0>$
amplitude:
\begin{equation}
<\pi^+(p_+)\pi^-(p_-)|\sum_{i=1}^3
c_iO_i|K^0(q)>=\frac{1}{4}\sqrt{2}F_{\pi}(c_1-c_2-c_3)(q^2-p^2_-)
\end{equation}
The combination $c_5O_5+c_6O_6$, after reordering of quarks in the spinor
and colour spaces, turns into
\begin{equation}
c_5O_5+c_6O_6=\frac{8}{9}\tilde c_5 \bar s(1-\gamma_5)q \cdot \bar
q(1+\gamma_5)d, \qquad \tilde c_5=c_5+\frac{3}{16} c_6.
\end{equation}
After bosonization
\begin{equation}
\begin{array}{lll}
c_5O_5 +c_6O_6=\frac{8}{9}  \tilde c_5 \frac{2F_{\pi}^2
m^4_{\pi}}{(m_u+m_d)^2}[(\hat \sigma -i\hat \pi)(\hat \sigma+i\hat
\pi)]^{P-odd}_{23} =\frac{8}{9}
i \tilde c_5 \frac{2F^2_{\pi}m^4_{\pi}}{(m_u+m_d)^2}[\sigma_{\pi^-}K^+ \\
- \pi^- \sigma_{K^+}  +\sqrt{\frac{3}{2}}(\sigma_8 K^0-\pi_8
\sigma_{K^0})- \sqrt{\frac{1}{2}}(\sigma_3 K^0-\pi_3 \sigma_{K^0})]
\end{array} \end{equation} A breakdown of the $SU(3)$ symmetry together
with mixing of $\sigma_0$ with gluonium originates a mixing between
$\sigma_0$ and $\sigma_8$, so that the physical states become of the forms
\begin{equation}
\sigma_{\eta'}=\sigma_0 \cos \theta_S + \sigma_8 \sin \theta_S, \quad
\sigma_{\eta} = -\sigma_0 \sin \theta_S + \sigma_8 \cos \theta_S.
\end{equation}
To find the $K\to \pi^+\pi^-$ amplitude generated by $O_5$ operator, it is
necessary to know the amplitudes of $K^0 \to \sigma_{K^+} \pi^-$  and
$\sigma_8 \to \pi^+ \pi^-$ transitions. They are determined by the
lagrangian \cite{14}, \cite{15}:
\begin{equation}
\begin{array}{ll}
L_{mesons}=\frac{1}{2}Tr\{\partial_{\mu}U \partial_{\mu} U^{\dag}
\}- cTr\{U U^{\dag} -A^2 t^2_0\}^2 \\  -c\xi(Tr\{U U^{\dag} -A^2
t^2_0\})^2+ \frac{F_{\pi}}{2\sqrt{2}}Tr\{M(U+U^{\dag} )\} + \Delta
L^{U_1}_{PS} \end{array} \end{equation}
This lagrangian is considered here as the phenomenological effective
lagrangian where the baryonic degrees of freedom are integrated out.
In addition, the masses and coupling constants are such that their
renormalization generated by mesonic loops is not needed. However, the
renormalization procedure does not remove the corrections which are the
finite functions of the external momenta. They have to be taken into
account and will be considered later.

The parameter $\xi$ in eq.(11) characterizes a degree of mixing between
$\sigma_0$ and gluonium. Such a mixing increases $ <2\pi;I=0|K^0>$
amplitude \cite{2}, but in present paper, devoted to investigation of FSI
role, we shall consider, for simplicity, the case $\xi=0$ in which $\sin
\theta_S=1/\sqrt{3}$. Then (see \cite{15})
\begin{equation}
\begin{array}{lll}
g_{\sigma_{\eta'} \pi\pi}=-(m^2_{\sigma_{\eta'}}-m^2{\pi})/F_{\pi}, \quad
g_{\sigma_{\eta} \pi\pi}=0,  \\ g{\sigma_{K^+} \bar K^0
\pi^-}=-(m^2_{\sigma_{\eta'}}-m^2_{\pi})(2R-1)/\sqrt{2}F_{\pi} \\
m^2_{\sigma_{\eta'}}-m^2_{\pi}=(m^2_K-m^2_{\pi})/[(R-1)(2R-1)], \\
m^2_{\sigma_K}-m^2_{\pi}=(m^2_{\sigma_{\eta'}}-m^2_{\pi})(2R-1)R
\end{array}
\end{equation}
where $R=F_K/F_{\pi}$. In our theory
\begin{equation}
R=1+(m^2_K-m^2_{\pi})/(m^2_{\sigma_{\eta'}}-m^2_{\pi})....
\end{equation} The amplitude generated by operators $O_{5,6}$ is \cite{2}:
\begin{equation} <\pi^+\pi^-|c_5O_5 +c_6 O_6|K^0(q)>=\frac{8}{9}
\frac{\sqrt{2}F_{\pi}m^4_{\pi}}{(m_u+m_d)^2}\tilde
c_5\left[\frac{m^2_{\sigma_{\eta'}}-m^2_{\pi}}{m^2_{\sigma_{\eta'}}-q^2}-
\frac{1}{R} \right].
\end{equation}
Neglecting the higher-order corrections, we obtain the total amplitude
\begin{equation}
\begin{array}{lll}
<\pi^+(p_+)\pi^-(p_-)|\sum \limits_{i=1}^6 c_iO_i|K^0(q)>= \\
\frac{1}{4}\sqrt{2}F_{\pi}\left[(c_1-c_2-c_3)(q^2-p^2_-) +\frac{32}{9}
\beta \tilde c_5 (m^2_K-m^2_{\pi}) \right]
\end{array}
\end{equation}
where
\begin{equation}
\beta=\frac{2m^4_{\pi}}{(m_u+m_d)^2 (m^2_{\sigma{\eta'}}-m^2_{\pi})}
\end{equation}
Such is the amplitude $<2\pi;I=0|K^0>$ produced by weak interaction and
incorporating the corrections produced by strong quark-gluon interactions
at short distances. Our next step is to find the changes arising due to
rescattering of pions at large distances where the real intermediate
hadrons play role. We begin from a consideration of the elastic $\pi\pi$
scattering itself.
\section{The elastic $\pi\pi$ scattering}
An amplitude of elastic $\pi\pi$ scattering depends on two variables:
initial energy of two pions and their scattering angle in c.m. system.
When this amplitude is expressed in terms of the phase shifts, the partial
amplitudes depending only on $s=(p_1+p_2)^2$ are the objects of
investigation. Let's the $S$-wave partial amplitude (the only one
participating in $K\to 2\pi$ decay ) calculated in the tree
approximation is
\begin{equation}
A^{tree}=f(s).
\end{equation}
To one-loop it turns into
\begin{equation}
A^{one-loop}=f(s)[1+\mbox{Re} \Pi_R(s)+i\mbox{Im} \Pi(s)]=f(s)[1+\mbox{Re}
\Pi_R(s) +i \frac{f(s)\sqrt{1-4m^2/s}}{16\pi}] \end{equation} where
$\mbox{Re} \Pi_R(s)$ means the finite part depending on external momenta
remaining after regularization of $\mbox{Re} \Pi(s)$.

The unitarization, representing the summing up of the chains built from
the loop diagrams corresponding to rescattering of pions, leads to
\begin{equation}
A^{unitar}= \frac{f(s)}{1-\mbox{Re}\Pi_R(s) -i\mbox{Im}\Pi_R(s)}=
 =\frac{f(s)}{1-\mbox{Re}\Pi_R(s)}\cdot
\frac{1}{1-i\frac{f(s)\sqrt{1-4m^2/s}}{16\pi(1-\mbox{Re} \Pi_R(s))}}
\end{equation}
Introducing the notation
\begin{equation}
\frac{f(s)\sqrt{1-4m^2/s}}{16\pi(1-\mbox{Re}\Pi_R(s))} =\tan \delta,
\end{equation}
we come to the known from the non-relativistic quantum mechanics
expression
of the scattering amplitude through the phase shifts:
\begin{equation}
A^{unitar}=
\frac{16\pi\sin \delta e^{i\delta}}{\sqrt{1-4m^2/s}}
\end{equation}
leading to the cross-section:
\begin{equation}
\sigma=\frac{4\pi\sin^2 \delta}{k^2}, \qquad k=\frac{\sqrt{s}}{2}
\sqrt{1-4m^2/s}
\end{equation}

From eq.(20) and representation of $A^{unitar}$ in the form (19)
\begin{equation}
A^{unitar}=\frac{f(s)}{1-\mbox{Re} \Pi_R(s)} \cos \delta e^{i\delta},
\end{equation}
one concludes that at $\mbox{Re} \Pi_R(s)<0$ (more exactly, $\mbox{Re}
\Pi_R(s)<1-\cos \delta$) FSI diminishes the absolute value of the initial
amplitude (17).

It should be noted in addition that according to eq.(20), $\mbox{Re}
\Pi_R(s)$ can be found using the data on the phase shifts and the
theoretical value of $f(s)$ obtained in one or another model of $\pi\pi$
interaction. And this is extremely important because the direct
calculation of the contribution from the closed  hadron loops in the
lowest order of perturbation theory does not allow to obtain the right
conclusion on a value of $\mbox{Re}\Pi_R (s)$. A reason is that such a
contribution turns out to be of order 1. In such case the
perturbation theory is not applicable.

In the literature devoted to study of the role of the resonances in
mesonic theory, a necessity of incorporation of $ \mbox{Re}\Pi_R$ in
analysis was not considered, as a rule \cite{16}. An exception are the
papers \cite{17} where $\mbox{Re} \Pi_R$ was calculated in the lowest
order of perturbation theory, which does not work in the case of strong
coupling, as it was mentioned above. Nevertheless, it is possible to obtain
the reliable estimate of $\mbox{Re} \Pi_R$. As the function $f(s)$ in the
chiral theory is proportional to $g^2_0$, the effect of $\mbox{Re}\Pi_R$
can be transfered into $g^2(s)$:
\begin{equation}
g^2_0/(1-\mbox{Re} \Pi_R(s)) \Longrightarrow g^2(s)=g^2_0 F(s),
\end{equation}
where $F(s)$ is the phenomenological form factor choosed so that the
theoretical results for the set of phase shifts $\delta^0_0, \delta^2_0,
\delta^1_1, \delta^0_2$ and $\delta^2_2$ should coincide with the
experimental data in some broad range of values of $s$. This
method, simplifying an analysis  of the effects of $\mbox{Re} \Pi_R(s)$
was used in \cite{18} where the Chiral Resonance Theory of $\pi\pi$
Scattering in the range $4m^2_{\pi}\le s \le 1\;\mbox{GeV}^2$ was
elaborated. It follows from \cite{18} that the main contribution into
$\mbox{Re} \Pi_R(s=m^2_K)$ is generated by $\sigma \pi \pi$ interaction
and
\begin{equation}
F(s=m^2_K)\approx \mbox{exp}(-(m^2_K-m^2_{\pi})/2\mbox{GeV}^2)=0.894.
\end{equation}
that is
\begin{equation}
\mbox{Re} \Pi_R(s=m^2_K) \approx -0.12.
\end{equation}
The sign of $\mbox{Re} \Pi_R(s)$ coincides with the one following from the
calculation in the lowest order of perturbation theory \cite{17}, but the
absolute value turns out to be considerably smaller.

Now, let us pass to study of the FSI influence on the amplitude of $K \to
2\pi$ decay.
\section{FSI in $K^0 \to 2\pi$ decay}
To one loop, the amplitude (15) transforms into
\begin{equation}
\begin{array}{llll}
<\pi^+ (p_+) \pi^-(p_-)|\sum \limits^6_{i=1} c_iO_i|K^0(q)>^{one-loop}=
\frac{F_{\pi}}{2\sqrt{2}} \{(c_1-c_2-c_3)[(q^2-p^2_-)+ \\
\frac{f(s)}{(2\pi)^4 i}\int\frac{(q^2-p^2)d^n
p}{[(p-q)^2-m^2_{\pi}][p^2-m^2_{\pi}]}+i\frac{f(s)}{16\pi}\sqrt{1-
4m^2_{\pi}/q^2}(q^2-p^2_-)] \\
+\frac{32}{9}\beta \tilde c_5 (m^2_K-m^2_{\pi})[1+
\frac{f(s)}{(2\pi)^4 i} \int \frac{d^n p}{[(p-q)^2
-m^2_{\pi}][p^2-m^2_{\pi}]}+ i\frac{f(s)}{16\pi}\sqrt{1-4m^2/q^2}]
\}
\end{array}
\end{equation}
Taking into account that the part of the first integral proportional to
finite function of the external momenta is equal to (see Appendix)
\begin{equation}
(q^2-m^2_{\pi})\frac{f(s)}{16\pi^2}
\sqrt{1-4m^2/q^2}\ln\frac{1-\sqrt{1-4m^2_{\pi}/q^2}}{1+\sqrt{1-
4m^2_{\pi}/q^2}}=(q^2- m^2_{\pi}) \mbox{Re}\Pi_R(q^2) \end{equation} and
the analogous part of the second integral is $$ \mbox{Re}\Pi_R(q^2) $$ and
accomplishing the unitarization according to the prescription (19) we come
to \begin{equation}
<\pi\pi;I=0|K^0(q^2=m^2_K)>=\frac{F_{\pi}}{2\sqrt{2}}(m^2_K-m^2_{\pi})[c_1-c_2
-c_3 +\frac{32}{9}\beta \tilde c_5] \frac{\cos \delta e^{i\delta}}{1-\mbox
{Re} \Pi_R(m^2_K)}
\end{equation}
Remembering that the integrals in eq.(27) calculated in the leading order
of perturbation theory do not give a reliable estimate of $\mbox{Re}
\Pi_R$ and using the estimate (26) together with $\delta^0_0 \approx
37^{\circ}$ \cite{18} we come to conclusion that FSI diminishes the tree
amplitude (15) by 30\%.

The analogous analysis of the influence of FSI on the amplitude
$<\pi\pi;I=2|K^0>$ leads to conclusion that FSI enlarges this amplitude by
5\%.
\section{Conclusion}
We did not find an enlargement of the amplitude $<\pi\pi;I=0|K^0>$ due to
final state interaction of pions. On the contrary, our analysis showed
that FSI diminishes this amplitude. The amplitude $ <\pi\pi;I=2|K^0>$ is
increased a little by FSI. So that, FSI is not at all the mechanism
bringing us nearer to explanation of the $\Delta I=1/2$ rule in $K\to
2\pi$decay. But our result, of course, does not mean that this rule can
not be explained.In particular, an increase of the coefficient $\tilde
c_5$ calculated in theory with large uncertainty (see \cite{1}) can
compensate the negative influence of FSI.
\vspace{2cm}
\section*{Appendix}
The loop integral
\begin{equation}
I^{(0)}=\frac{1}{(2\pi)^4}\int \frac{d^n p}{[(p-q)^2-m^2][p^2-m^2]}
\end{equation}
calculated using the t'Hooft-Veltman dimensional regularization \cite{19}
is equal to
\begin{equation}
I^{(0)}=\frac{i}{16\pi^2} \left( \ln \frac{M_0^2}{m^2}
+2+\sqrt{1-4m^2/q^2} \ln \frac{1-\sqrt{1-4m^2/q^2}}{1+\sqrt{1-4m^2/q^2}}
\right)
\end{equation}
where
$$
\ln\frac{M^2_0}{\mu^2}=1/\epsilon -\gamma_E +\ln(4\pi), \qquad \epsilon
\to 0
$$
and $\mu$ is some arbitrary mass disappiaring in the above expression for
$I^{(0)}$.
After regularization removing from $I^{(0)}$ the terms independent on the
external momenta we obtain
\begin{equation}
I^{(0)}_R=
\frac{i}{16\pi^2}\sqrt{1-4m^2/q^2} \ln
\frac{1-\sqrt{1-4m^2/q^2}}{1+\sqrt{1-4m^2/q^2}}
\end{equation}
The first loop integral in eq.(27) is
\begin{equation}
\begin{array}{lll}
I^{(1)}= \frac{1}{(2\pi)^4}\int \frac{(q^2-p^2) d^n
p}{[(p-q)^2-m^2][p^2-m^2]}=  \\
= \{q^2
I^{(0)}-i\frac{m^2}{16\pi^2}\left(2\ln\frac{M^2_0}{m^2}+3+\sqrt{1-4m^2/q^2}\ln
\frac{1-\sqrt{1-4m^2/q^2}}{1+\sqrt{1-4m^2/q^2}} \right) \}
\end{array}
\end{equation}
After regularization it converts into
\begin{equation}
I^{(1)}_R=(q^2-m^2) I^{(0)}_R.
\end{equation}
Just this part of $I^{(1)}$ turns out to be proportional on mass shell to
the combination $(m^2_K-m^2_{\pi})$ that vanishes in the limit of exact
SU(3) symmetry in accordance with Cabibbo-Gell-Mann theorem for $K \to
2\pi$ amplitude \cite{20}.

For  $ \Pi_R(s)$ in eq.(18), it can be obtained in the lowest order of
perturbation theory (see \cite{17}):
\begin{equation}
\begin{array}{llll}
  \Pi_R(s)=\frac{f(s)}{16\pi^2}\sqrt{1-4m^2/s}\ln
\frac{1-\sqrt{1-4m^2/s}}{1+\sqrt{1-4m^2/s}} +
i\frac{f(s)}{16\pi}\sqrt{1-4m^2/s},\quad s>4m^2 \\
\Pi_R(s)=-\frac{f(s)}{16\pi}|\sqrt{1-4m^2/s}|\cdot(1-\frac{2}{\pi}\arctan|\sqrt{1
-4m^2/s}|),\quad s<4m^2.
\end{array}
\end{equation}
But outside the perturbation theory, a value of $\mbox{Re}\Pi_R(s)$
depends on the used model of $\pi\pi$ interaction. In our case this value
is given by eq.(26).

\end{document}